# Collection of micromirror-modulated light in the single-pixel broadband hyperspectral microscope

Lukáš Klein, Karel Žídek[*]

*Regional Centre for Special Optics and Optoelectronic Systems (TOPTEC), Institute of Plasma Physics, Academy of Sciences of the Czech Republic, Za Slovankou 1782/3, 182 00 Prague 8, Czech Republic.*

[*]*Author to whom correspondence should be addressed: zidek@ipp.cas.cz*

## Abstract

Digital micromirror device (DMD) serves in a significant part of computational optical setups as a means of encoding an image by the desired pattern. The most prominent is its usage in the so-called single-pixel camera experiment. This experiment often requires an efficient and homogenous collection of light from a relatively large chip on a small area of an optical fiber or spectrometer slit. Moreover, this effort is complicated by the fact that the DMD acts as a diffractive element, which causes severe spectral inhomogeneities in the light collection. We studied the effect of light diffraction via a whiskbroom hyperspectral camera in a broad spectral range. Based on the knowledge, we designed a variety of different approaches to light collection. We mapped the efficiency and spectral homogeneity of each of the configuration - namely its ability to couple the light into commercially available fiber spectrometers working in the visible and IR range (up to 2500 nm). We found the integrating spheres to provide homogeneous light collection, which, however, suffers from very low efficiency. The best compromise between the performance parameters was provided by a combination of an engineered diffuser with an off-axis parabolic mirror. We used this configuration to create a computational microscope able to carry out hyperspectral imaging of a sample in a broad spectral range (400-2500 nm). We see such a setup as an ideal tool to carry out spectrally-resolved transmission microscopy in a broad spectral range.

## I. Introduction

Digital micromirror device (DMD) has become a commonly used means of modulating light in a variety of applications and scientific experiments, including projectors, UV lithography, microscopy, and many others.[1] In particular, in the field of compressed sensing, the DMD has become a dominating device, which is commonly used to attain well defined binary patterns for signal encoding.[2,3] Compressed sensing is a novel approach to data acquisition, where a randomly or pseudorandomly modulated signal can be reconstructed back from a reduced dataset.[2,4,5] The most common experiment is the so-called single-pixel camera (SPC), where the image of interest is superimposed on a random mask, and the total intensity of the masked image is measured by a single-pixel detector, such as a photodiode or a photomultiplier.[6–8] A small fraction of measurements – approx. 20% compared to the number of pixels – can be used to retrieve back the image.[2]

One of the issues, which is often acquainted in the SPC experiments, is to efficiently and evenly collect light from a relatively large DMD (sized several millimeters) onto a small area of a detector, optical fiber, or a spectrograph slit (typically well below 1 mm). Since the micromirror elements feature size comparable to the light wavelength, the light diffraction inevitably occurs and becomes one of the biggest problems, when broadband white light is collected. This issue becomes progressively more pronounced in the infrared region. A variety of solutions can be traced in the literature, such as downscaling the DMD image by a 4f lens imaging and building a custom-made spectrograph,[9] or sacrificing light intensity on the expenses of an even light collection. Nevertheless, there is no generally accepted solution to the problem, and articles seldom address the issue of diffraction on the encoding mask or DMD.[10,11]

In this article, our goal was to set up an SPC experiment, where a large variety of light collection configurations can be tested. For the same conditions and collecting optical fiber, we analyzed characteristics of six different configurations with respect to the total and spectral homogeneity of the collected light, and also the overall efficiency of the light collection. In order to understand and describe the properties of the DMD-reflected light, we also acquired a hyperspectral image of the DMD-diffracted white light.

Based on the gained knowledge, we chose the best configuration for a hyperspectral SPC and we carried out a set of experiments demonstrating the possibility to use the setup for a hyperspectral imaging in the standard silicon-detector range (200-950 nm), as well as the option to get a simple IR hyperspectral camera working with a moderate image quality (800-2500 nm) by using a compact fiber NIR spectrometer.

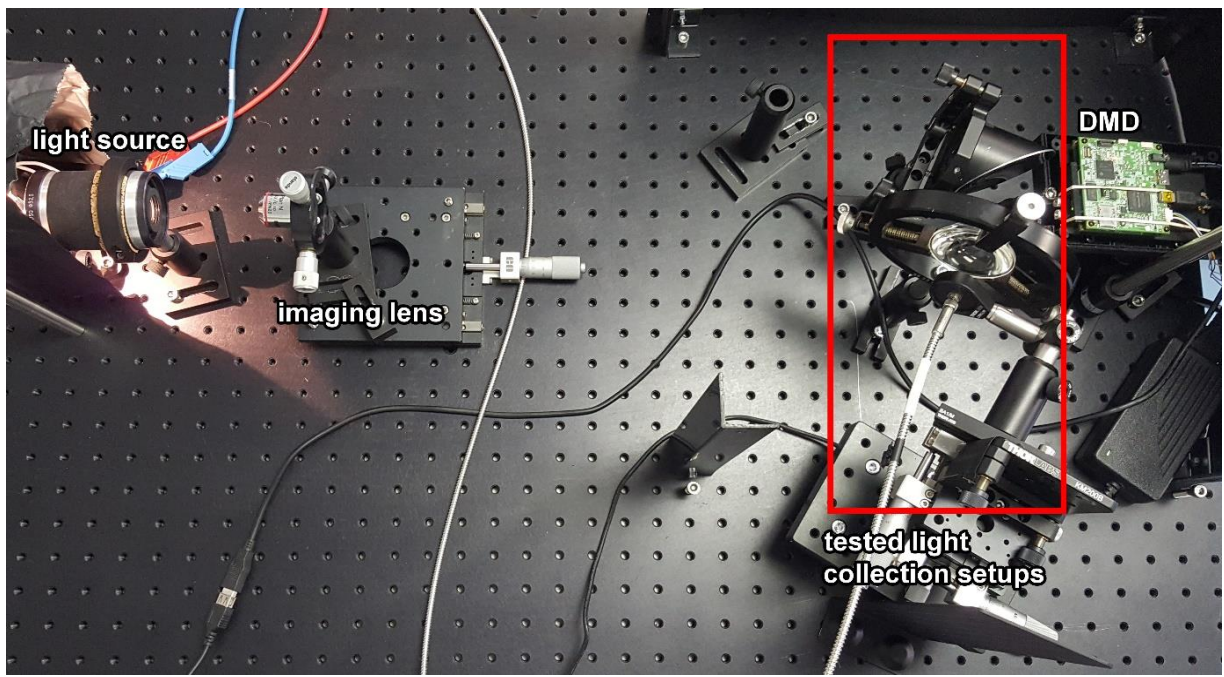

**Figure 1: Configuration of the testing setup, where the light was imaged by a microscopic imaging objective onto a DMD. The light collection arrangement (denoted by red frame) was altered for each tested configuration.**

## II. Methodology

The used experimental setup is depicted in Figure 1. In all the presented measurements, each configuration used a modified light collection part (denoted by a red line), whereas the remaining parts of the setup stayed unchanged.

As a light source, we used a standard tungsten halogen lamp (12 V, 1.7 A) due to its broad continuous spectrum. The lamp was powered from a stabilized laboratory power supply. Light from the lamp illuminated the focal plane of a microscopic lens (Olympus Plan Achromat 4x, NA = 0.1). The focal plane was the sample space in the SPC experiment. The light was imaged by the microscopic lens onto a DMD (DLP3000 chip, Texas Instruments). Subsequently, a variety of optical configurations was used to collect the light. For the sake of clarity, each specific configuration is described in the following section, together with the individual results. The DMD chip was slightly titled in order to reflect the light upwards. The tilt allowed us to place the necessary optics close to the DMD chip. At the same time, the tilt has been set to the minimum value in order to attain a sharp image of a sample on the chip.

Hyperspectral image of the diffraction pattern from DMD was acquired by the so-called whiskbroom technique. An optical fiber 1.6 mm in diameter with a high numerical aperture of 0.5 was used to collect light from a plane 80 mm from the DMD chip. By using a motorized XY translational stage, we scanned the area of 25 by 25 mm.

The light was analyzed by using compact fiber spectrometers Ocean Optics, Flame for the visible part of the spectrum (VIS, 200-1000 nm), NIR-quest 256 for the IR spectrum (800-2500 nm).

The light collection homogeneity was analyzed by dividing the DMD micromirrors into 8×8 sections, where the micromirrors in a selected section were all set to 0 (reflection in the detector direction), and the remaining part was set to 1, i.e., it was not detected. By scanning the sections one-by-one, we obtained the necessary information. To gather the spectrum all mirrors were tilted towards the detector and the whole area was measured at once. The light was collected through a fiber 1.6 mm in diameter, NA 0.5 to the compact fiber spectrometers (Flame and NIR-Quest, both Ocean Optics). SPC experiments employed the same configuration, where the DMD chip was used to create a random pattern, as we describe later.

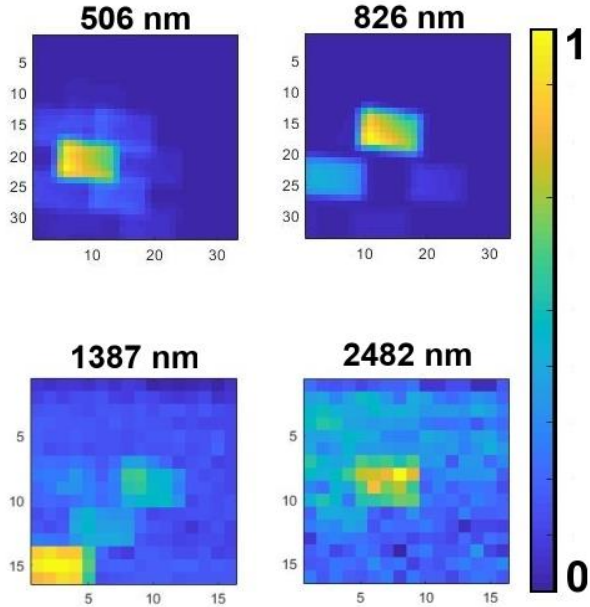

Figure 2: Diffraction patterns from the used DMD chip, which was acquired for selected wavelengths in the distance of 80 mm. The entire field of view corresponds to 25x25mm square. Visible spectral region (upper panels) were scanned with 32x32 resolution. IR spectral region (lower panels) with 16x16 stepping.

## III. Results and discussion

Since a DMD chip consists of a 2D array of mirrors sized several micrometers (7.637 μm in our case), light diffraction has to occur inevitably. In order to study its properties, we have created a hyperspectral image of the diffraction patterns in the range of 300-2500 nm. As expected, we observed that the effect becomes more pronounced in the IR region, with the diffraction maxima increasing their distance (see Figure 2). Since the micromirrors are arranged in the diamond-like fashion, the diffraction maxima occur in the diagonal directions. Because the DMD chip acts, to some extent, as a blazed grating, the diffraction efficiency to each mode is significantly spectrally dependent. This effect is visible in the four patterns in Figure 2, where the most intense diffraction order varies from case to case. The variation is the primary source, which causes large spectral inhomogeneities in the light collection from DMD, as we will describe later.

The observed diffracted patterns provide us with the information that a spectrally homogeneous light collection requires gathering photons from a wide-angle, which includes the entire zero and first diffraction modes. This becomes progressively more demanding for the IR light. For example, in the section depicting diffraction for 826 nm the edge of the first mode is nearly 13 mm away from the center of the zero mode. For this wavelength, the combination of zero and first modes are reflected from DMD into a solid angle of 0.16 steradian.

Based on the attained knowledge, we proposed five different configurations for light collection, which are schematically depicted in the left column of Figure 3. For each setup, we measured light collected from 8x8 divisions of the DMD chip and analyzed for each section the light spectrum. The selected benchmark, which we use to compare the configurations, was the ability to couple light into an optical fiber 1.6 mm in diameter, NA of 0.5 The values of the total intensity of light (spectrally integrated) for each section were used to create homogeneity maps (middle column in Figure 3). Finally, we also measured the spectrum of light reflected from the entire DMD chip (right columns in Figure 3). Although the spectra of the DMD-reflected light slightly varied for each DMD section, the total spectrum in the right column gives a very good image of the spectral homogeneity of each configuration.

We illuminated the DMD with a white lamp, which forms a uniform spot in the sample plane with several millimeters in diameter. Based on the previous experiments, we know that with the microscope lens, the area of approx. 600x400 microns in the sample plane covers the DMD chip. Therefore, for the sake of our measurements, we can assume that the DMD is illuminated with a nearly homogeneous intensity. In all cases, the measurement configuration was, prior to the experiment, optimized in order to attain the most efficient overall light collection efficiency.

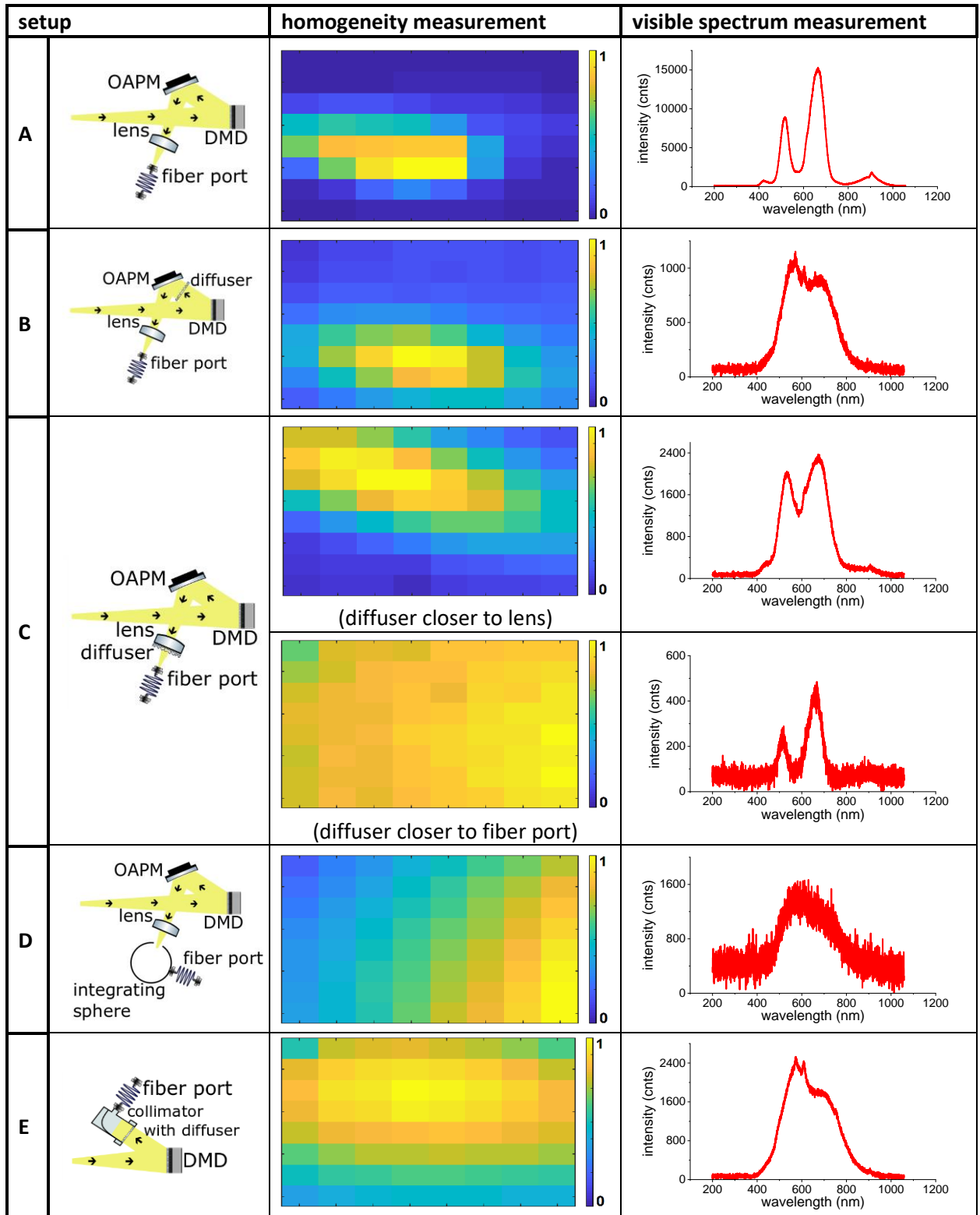

**Figure 3: Overall and spectral homogeneity of the collected light from each tested configuration.** Left panels: scheme of the tested configuration; middle panels: spectrally-integrated homogeneity of the collected light, 8x8 division of the DMD chip, all maps are normalized to their maximum intensity; right panels: spectral homogeneity of light integrated from the whole DMD chip.

**Setup A** represented the most straightforward approach to the problem, the off-axis parabolic mirror (Thorlabs MPD249-G01) together with a lens (plano-convex, f = 50 mm) were used to create a scaled-down image of the DMD chip. The homogeneity map shows that there was practically no information gathered from the edges of the DMD, and there were massive falls of intensity for specific wavelengths.

In **setup B,** we added a diffuser, which was set in the DMD proximity. The added diffuser improved both the spatial and spectral homogeneity at the cost of 15-times lower overall intensity, as seen by the maximum values in both graphs. Moreover, the outer parts of the DMD chip were collected very inefficiently.

In s**etup C,** we placed the diffuser between the lens and fiber port. Here we tested two options by setting the diffuser in the lens and fiber port proximity. As in the previous result, we observe that this approach can be used to homogeneously gather narrow band light when the diffusor is close to the fiber port. For the broadband light, this option, however, cannot be used.

In **setup D,** we used an integrating sphere attached to the fiber port. Because the commercially available integrating spheres were unnecessarily large for our application, which implies a significant intensity loss, we fabricated an integrating sphere optimized for this experiment. Based on the DMD image produced by an off-axis parabolic mirror, we set the input port diameter to 5 mm. The output port was set by the optical fiber dimensions (SMA connector). By using the BaSO4 layer deposited on a sphere with the inner diameter of 16 mm, we were able to improve 4-fold results, compared to the standard 2-inch integrating sphere. This configuration provided the most continuous spectrum out of all, while only receiving a tiny fraction of overall intensity. When the light collection was optimized for maximum overall efficiency, the homogeneity showed a linear decrease from one side to another. The decrease is likely caused by the fact that the light detection is asymmetric, as the collecting optics have to be placed so that the DMD chip is not blocked. Using an integrating sphere is the most suitable approach for the spectrally broad light, where the photon budget of the detection is not a significant issue, such as time-correlated single-photon counting setups.

The best compromise between the collection inhomogeneity, the overall efficiency, and the spectral homogeneity was attained by using **setup E.** Here we used an off-axis parabolic mirror (a commercially available parabolic collimator Thorlabs RC08SMA-P01) with an engineered circle pattern diffuser installed at its entry port and the fiber connected to the collimator. Owing to its small dimensions, the collimator was mounted close to the DMD (30 mm), highly improving the angle of light. As one can see, the spectral homogeneity is on the level of an integrating sphere, while the out parts of the DMD are collected with an efficiency exceeding 50 % compared to the central region.

The setup E was then used to test the **compressive hyperspectral imaging**. Since the introduction into the compressive imaging, including single-pixel camera, is well described in previously published literature[12,13] and a detailed description of the used measurement setup and reconstruction routines is presented in our previous work,[7] we direct reader to these articles.

The possibility to acquire a visible (VIS) or infrared (IR) hyperspectral information is provided by detecting the total light intensity by a standard fiber spectrometer using a Si-based detection (200-1100 nm) and InGaAs-based detector (800-2500 nm). We tested the setup in both modes.

For the visible spectrum, we used a combination of LEGO bricks of different colors as a testing sample (see Figure 1A, bright central spot), providing a set of materials with various VIS spectral transmittance. The image was reconstructed using the number of measurements $M = 1296$. This corresponded to 25 % of the number of pixels of the 72x72 pixel image $N = 5184$. An intense light transmitted between the neighboring bricks forms the main cross feature of the picture. However, the attained light intensity for red, green, and blue spectral regions reveals that the reconstructed light intensity transmitted through a brick follows the visually perceived colors.

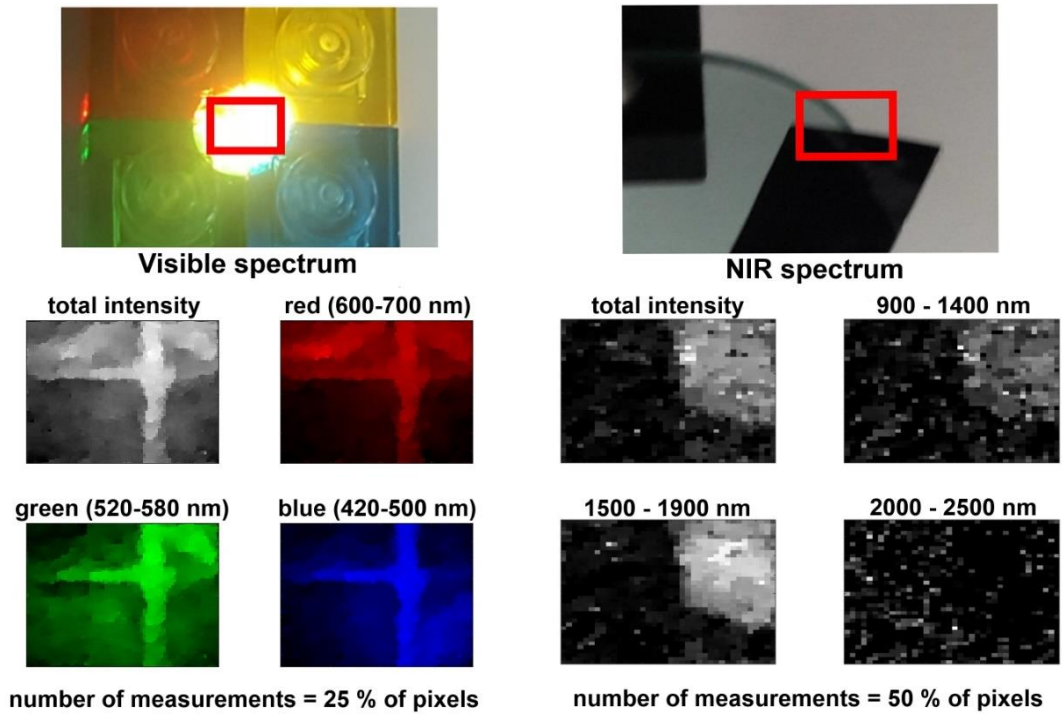

**Figure 4:** Compressive hyperspectral imaging using setup E in the 400-900 nm spectral range (VIS, left-hand side) and 800-2500 nm spectral range (NIR, right-hand side). Upper pictures show the used samples, and the red rectangle denotes the imaged area. Lower images represent reconstructed images for the total light intensity (left upper corner) and for three selected spectral bands (other images). VIS hyperspectral images feature the resolution of $N$ = 72x72 = 5184 pixels, number of measurements $M$ = 1296. NIR hyperspectral images: N =36x36 = 1296; M = 648.

The IR hyperspectral imaging was tested on a sample created by combining KG5 SCHOTT glass (transparent for visible light, opaque to NIR) and a polished silicon wafer (vice versa), while also leaving a part of the image unblocked. Here, the ratio $M/N$ = 648/1296 was set to 50% in order to compensate for a substantial dark current noise, which is typical for the IR detection. Due to the noise and high losses caused by the IR light diffraction, we could only clearly resolve the borderline between the uncovered and the covered area – see the bright area in the right upper corner. Nevertheless, the difference between KG5 and silicon wafer was not apparent in the hyperspectral results. Only in the total (whole NIR spectrum) reconstruction, we can see that the area in the right lower corner (Si wafer position) is slightly „brighter" compared to the left-hand side. Finally, the combination of water vapor absorption and a decreased detector sensitivity causes the lack of any signal in the 2000-2500 nm range (lower right panel).

We can, therefore, state that the presented setup can be used for standard hyperspectral imaging in the VIS range, while the IR measurement provides only a crude image and require a stronger light source for reasonable imaging.

## IV. Conclusions

In summary, we tested a number of setups in order to evaluate their potential of serving to collect light from a DMD chip into an optical fiber, spectrometer chip, or a small area detector in general. For versatile use, the combination of an engineered diffuser with a small off-axis parabolic mirror turned out to be the most effective. We have demonstrated that by using this configuration, we can build a hyperspectral camera employing commercially available fiber spectrometers.

For the measurements involving extremely high intensities of light or requiring a low light intensity, such as time-correlated single-photon counting, using the integrating sphere becomes the more viable option. This approach outperforms all other methods in the spectral homogeneity of the collected light. Moreover, since a DMD reflects light on both sides from the chip, we can combine the two options into a single setup and use the appropriate option for each experiment.

## Acknowledgment

We gratefully acknowledge the financial support of The Czech Academy of Sciences (ERC-CZ/AV-B, project Random-phase Ultrafast Spectroscopy; AV21 programme); Grant Agency of the Czech Republic (Project 17-26284Y); Ministry of Education, Youth and Sports („Partnership for Excellence in Superprecise Optics", Reg. No. CZ.02.1.01/0.0/0.0/16_026/0008390)